\documentclass[]{aa} 

\usepackage{graphicx}
\usepackage{txfonts}
\usepackage[colorlinks=true,linkcolor=blue,urlcolor=blue,citecolor=blue]{hyperref}
\usepackage{natbib}
\usepackage{morefloats}
\usepackage{epstopdf}
\bibpunct{(}{)}{;}{a}{}{,} 

\begin{document} 

\title{The effects of the stellar wind and orbital motion\\ 
on the jets of high-mass microquasars}

\author{ V. Bosch-Ramon\inst{1} \and
M.V. Barkov \inst{2}
}

\institute{Departament d'Astronomia i Meteorologia, Institut de Ci\`ences del Cosmos (ICCUB),
Universitat de Barcelona (IEEC-UB), Mart\'{\i} i Franqu\`es 1,
E-08028 Barcelona, Spain
\and
Astrophysical Big Bang Laboratory, RIKEN, 2-1 Hirosawa, Wako, Saitama 351-0198, Japan
}

\offprints{V. Bosch-Ramon, \email{vbosch@am.ub.es}}

   \date{Received - ; accepted -}
 
  \abstract
{High-mass microquasar jets propagate under the effect of the wind from the companion star, and the orbital motion of the binary system. The stellar wind and the orbit may be dominant factors determining the jet properties beyond the binary scales.}
{An analytical study is performed to characterize the effects of the stellar wind and the orbital motion on the jet properties.}  
{Accounting for the wind thrust transferred to the jet, we derive analytical estimates to characterize the jet evolution under the impact of the stellar wind. We include the Coriolis force effect, induced by orbital motion and enhanced by the wind presence. Large-scale evolution of the jet is sketched accounting for wind-to-jet thrust transfer, total energy conservation, and wind-jet flow mixing.}
{If the angle of the wind-induced jet bending is larger than its half-opening angle, the following is expected: (i) a strong recollimation shock; (ii) bending against orbital motion, caused by Coriolis forces and enhanced by the wind presence; and (iii) non-ballistic helical propagation further away. Even if disrupted, the jet can reaccelerate due to ambient pressure gradients, but wind entrainment can weaken this acceleration. On large scales, the opening angle of the helical structure is determined by the wind-jet thrust relation, and the wind-loaded jet flow can be rather slow.}
{The impact of stellar winds on high-mass microquasar jets can yield non-ballistic helical jet trajectories, jet partial disruption and wind mixing, shocks, and possibly non-thermal emission. Among several observational diagnostics at different bands, the radio morphology on milliarcsecond scales can be particularly insightful regarding the wind-jet interaction.} 

   \maketitle

\section{Introduction}\label{intro}

The jets of high-mass microquasars propagate through the wind of the companion star. This wind can strongly 
affect the jet hydrodynamics within the binary and farther away. This has been studied in a number of papers, mostly focusing on
dynamics, but sometimes also  considering the potential high-energy emission from the interaction, and in some cases accounting
for wind inhomogeneities \citep[e.g.][]{rom03,ore05,per08,owo09,ara09,per10,per12,bk12,yoo15,zdz15}. 

In addition to the role of wind thrust intercepting the jet, orbital motion can also have a dynamical influence on the jet on scales 
larger than those of the binary, inducing a helical pattern in the jet \citep[e.g., see the discussion in][]{bos13}. 
The strength of this influence should be determined by the wind-jet thrust relation: the stronger the jet bending due 
to wind impact is, the closer to the binary the Coriolis force should affect jet propagation. A similar effect is expected 
to take place in binary systems hosting a massive star and a non-accreting pulsar, in which the stellar and the pulsar 
wind collide, and the shocked flows form a strongly non-ballistic spiral structure \citep{bos11,bos12,bed13,bos15,bar15}. 

In this work, we characterize through analytical estimates the way the stellar wind and orbital motion affect the jet in a
high-mass microquasar.  The results can be easily scaled to derive conclusions for relevant sources, like for instance Cyg~X-1 or
Cyg~X-3,  two high-mass microquasars suggested to be candidates of wind-jet interaction to explain observations in
different energy bands \citep[e.g.][and see also Sect.~\ref{obssig}]{alb07,rom10,dub10,vil13}.

\section{Wind-jet interaction}

\begin{figure}
\centering
\resizebox{\hsize}{!}{\includegraphics{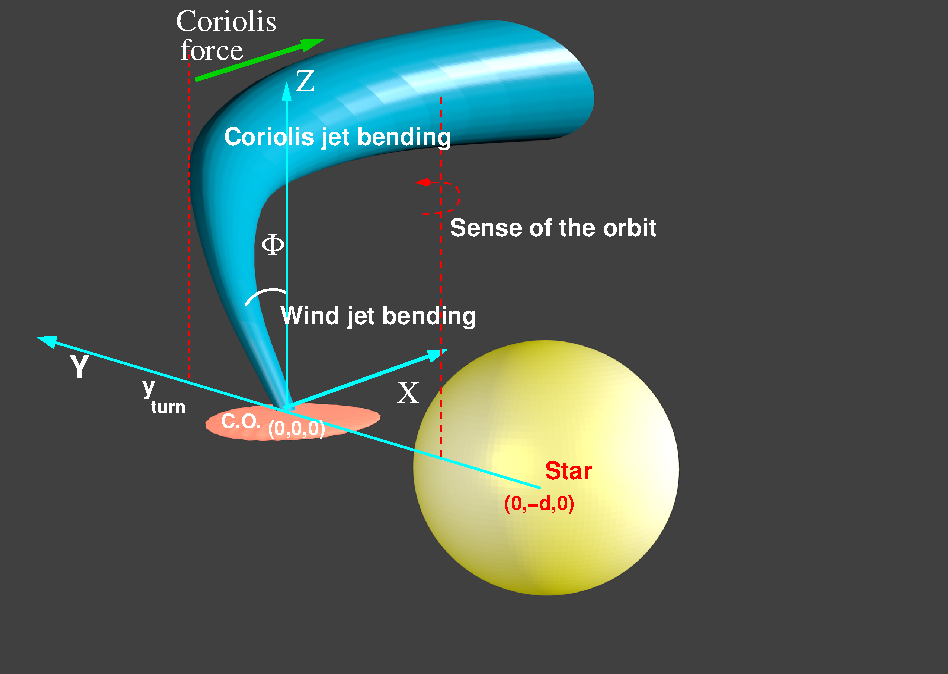}}
\caption{Sketch of the scenario studied in this work on small and middle scales; the main ingredients (star, stellar wind, jet), and physical processes and effects (wind-jet bending, Coriolis effect-jet bending) are shown.}
\label{f1}
\end{figure}

\begin{figure}
\centering
\resizebox{\hsize}{!}{\includegraphics{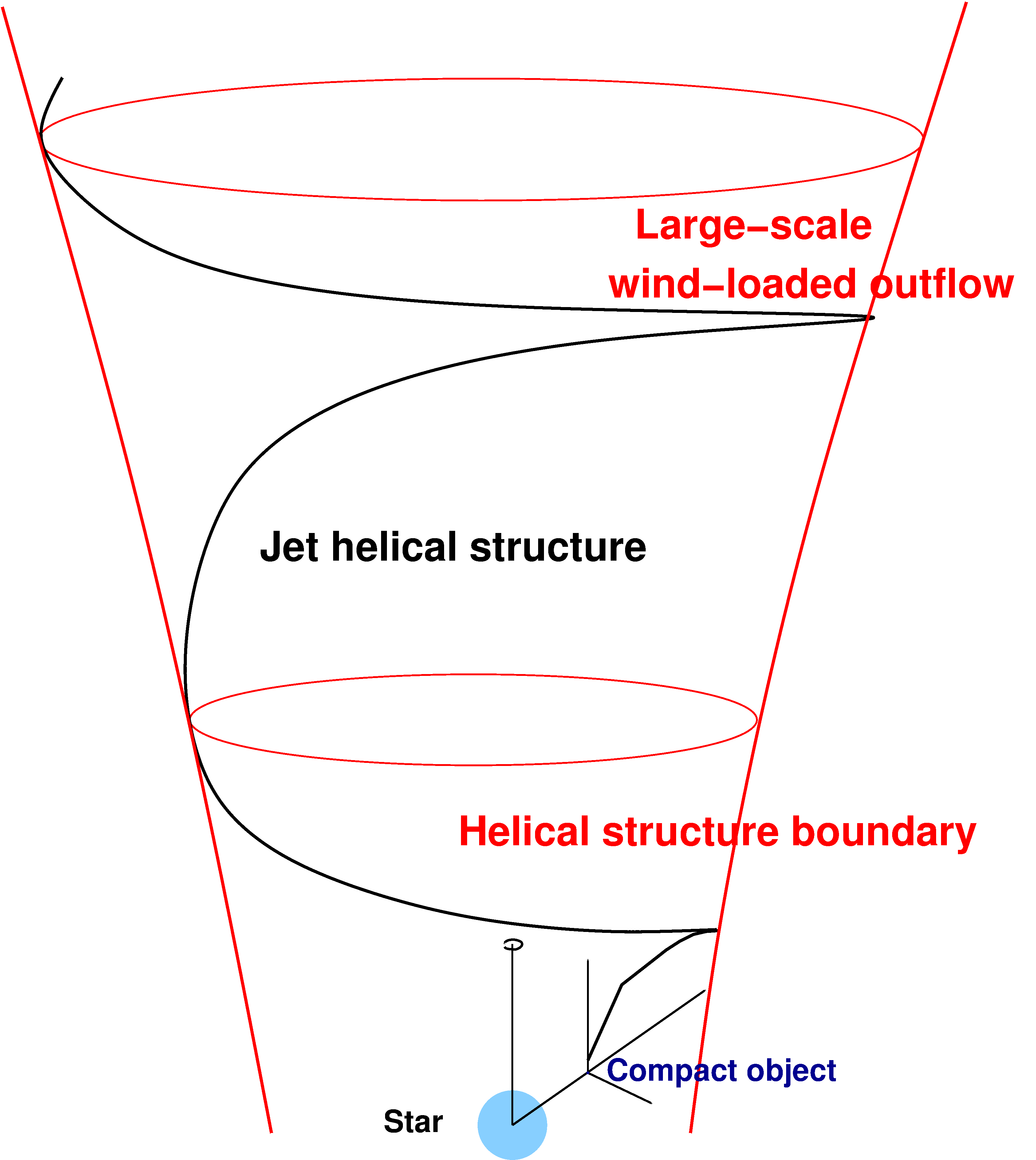}}
\caption{Sketch of the scenario studied in this work on the large scales; the main physical processes and effects (helical jet,  wind-jet outflow) are shown.}
\label{f2}
\end{figure}

Let us first set the scenario in which the wind-jet interaction problem is considered. The scenario consists of
a massive star with a strong and fast stellar wind, and a compact object, a black hole or an
accreting neutron star, from which a jet is launched. Typical values for the mass-loss rates and wind velocities
in OB stars are $\dot{M}_{\rm w}\sim 10^{-9}-10^{-5}\,$M$_\odot$~yr$^{-1}$ and $\varv_{\rm w}\sim 2\times 10^8$~cm~s$^{-1}$,
respectively \citep[e.g.][]{mui12,krt14}. Figures~\ref{f1} and \ref{f2} show a sketch of the scenario studied
here, with the different ingredients, physical processes and effects, and spatial scales discussed in this work.
As seen in the figures, the location of the star and the compact object are $(0,-d,0)$ and $(0,0,0)$, respectively. The stellar wind is assumed to be spherically symmetric, centred in the star, and of constant velocity. The initial jet direction is the $z$-axis (i.e. perpendicular to the orbital plane, which may not necessarily be the case), the $y$-axis goes first through the star and then through the compact object location, and the $x$-axis direction corresponds to $\bf{e}_x=\bf{e}_y\times \bf{e}_z$, pointing in the opposite direction to the sense of the orbit. 

It has been typically considered (though not jet proven) that persistent, low-hard state, microquasar jets are mildly relativistic at most \citep[e.g.][]{fen04}. Nevertheless, unlike some of the previous studies, which assumed a non-relativistic jet \citep{yoo15,zdz15}, we consider here that the jet could have a relativistic bulk velocity. This is justified by the fact that even modest Lorentz factors can have a significant impact when computing the jet lateral momentum flux ($f_{\rm jl}$), and indirectly other physical quantities. It is also worth noting that for a compact object mass of $M_{\rm co}\sim 3$~M$_\odot$, a distance of $z\sim 10^{11}-10^{12}$~cm means $\sim 10^5-10^6\,R_{\rm Sch}$, $R_{\rm Sch}=2GM_{\rm co}/c^2$ being the Schwarzschild radius if the compact object is a black hole. For a supermassive black hole of $10^8\,$M$_\odot$ hosted by an active galactic nuclei, this distance would correspond to $z\sim 1-10$~pc, typical scales on which the blazar emission, strongly affected by relativistic beaming, is produced \citep[e.g.][]{nal14,dot15}.

The convention $Q_x=(Q/10^x)$ is adopted all through the paper, with $Q$ in cgs units unless otherwise stated.

\subsection{Stellar wind impact: recollimation shock}

At some distance $z_{\rm i}$ from the compact object, the wind momentum flux can become larger than the lateral momentum flux of the jet, and stop the jet expansion towards the star \citep[e.g.][]{per08}. 

The distance $z_i$ depends on the initial jet half-opening angle, $\theta_{\rm j}$, the total stellar wind momentum rate
$\dot{P}_{\rm w}=\dot{M}_{\rm w}\varv_{\rm w}$, and the jet lateral momentum flux:
\begin{equation}
f_{\rm jl}=\rho_{\rm j}(\theta_{\rm j} u)^2=\frac{L_{\rm j}(\theta_{\rm j} u)^2}{\pi (\theta_{\rm j} z_i)^2
\gamma_{\rm j}(\gamma_{\rm j}-1)c^2u}=\frac{L_{\rm j}\beta_{\rm j}}{\pi z_i^2\gamma_{\rm j}(\gamma_{\rm j}-1)c}\,, 
\end{equation}
with $L_{\rm j}$, $\rho_{\rm j}$, $u$, and $\gamma_{\rm j}$ being the jet power (without accounting for its mass-rest energy), density, velocity, and Lorentz factor, and $\beta_{\rm j}=u/c$. 
The jet lateral momentum flux can also be written as a function of the jet total thrust,
\begin{equation}
\dot{P}_{\rm j}=\frac{L_{\rm j}}{c}\frac{\gamma_{\rm j}\beta_{\rm j}}{(\gamma_{\rm j}-1)}\,,
\end{equation}
as:
\begin{equation}
f_{\rm jl}=\frac{\dot{P}_{\rm j}}{\pi\,z_{\rm i}^2\gamma_{\rm j}^2}\,. 
\end{equation}
Before interacting, the wind propagation and jet expansion are assumed to be supersonic and non-relativistic, whereas as mentioned the jet propagation 
can be relativistic. Given the wind and jet geometries, the most intense wind-jet lateral interaction takes place within 
the binary system, on jet scales $z\lesssim d$. 

The value of $z_{\rm i}$ can be found from wind and jet-momentum flux balance in the $\bf{e}_y$-direction:
\begin{equation}
f_{\rm jl}=\frac{\dot{P}_{\rm j}}{\pi\,z_{\rm i}^2\gamma_{\rm j}^2}=\frac{\dot{P}_{\rm w}}{4\pi\,d^2}\left(\frac{d^2}{d^2+z_i^2}\right)^2\,{\rm .}
\label{balwj}
\end{equation}
There is a restriction on Eq.~(\ref{balwj}) to obtain a physical value for $z_i$ \citep[see][for the non-relativistic case]{zdz15}:
\begin{equation}
\left(\frac{4}{\gamma_{\rm j}}\right)\left(\frac{\dot{P}_{\rm j}}{\dot{P}_{\rm w}}\right)^{1/2}<1\,{\rm .} 
\label{m1}
\end{equation}
If the condition given in Eq.~(\ref{m1}) is fulfilled and the jet supersonic expansion is stopped by the wind, the jet suffers a recollimation shock \citep[e.g.][]{kf97,per08,bl09,per10,zdz15}. 
Otherwise, it implies that the jet is so powerful that $f_{\rm jl}$ is not balanced by the stellar wind at any point, which is the 
case for jet powers:
\begin{equation}
L_{\rm j}>\frac{\dot{P}_{\rm w}c^2\gamma_{\rm j}(\gamma_{\rm j}-1)}{16u}
\approx 
7\times 10^{36}\frac{\dot{M}_{\rm w,-7}\varv_{\rm w,8.5}\gamma_{\rm j,0.3}(\gamma_{\rm j,0.3}-1)}{\beta_{\rm j}}\,{\rm erg~s}^{-1}\,,
\label{recL}
\end{equation}
with $\dot{M}_{\rm w}$ units being M$_\odot$~yr$^{-1}$, and normalizing $\dot{M}_{\rm w}$ and $\varv_{\rm w}$ to values typical for high-mass stars.

\subsection{Stellar wind impact: jet bending}

To calculate the bending angle $\Phi$ of the jet due to wind impact, we introduce a non-dimensional parameter, $\chi_{\rm j}$, which corresponds to the ratio of jet-intercepting wind and jet thrusts for a conical jet perpendicular to the orbit:
\begin{equation}
\chi_{\rm j}=\frac{ \theta_{\rm j}\dot{P}_{\rm w}}{4\pi \dot{P}_{\rm j}}=\frac{\theta_{\rm j}\dot{M}_{\rm w}\varv_{\rm w}(\gamma_{\rm j}-1) c}{4\pi\gamma_{\rm j}\beta_{\rm j} L_{\rm j} }\approx 
0.02\frac{\theta_{\rm j,0.1}\dot{M}_{\rm w,-7}\varv_{\rm w,8.5}(\gamma_{\rm j,0.3}-1)}{\gamma_{\rm j,0.3}\beta_{\rm j}L_{\rm j,37}}\,.
\label{eq:Bj}
\end{equation}
The angle $\Phi$ can be then computed numerically in a self-consistent way, i.e. accounting for jet bending. The system of equations solved is presented in the Appendix. An approximate solution to these equations is (see Eq.~\ref{asimp}):
\begin{equation}
\Phi=\frac{\pi^2 \chi_{\rm j}}{2\pi \chi_{\rm j} +4\chi_{\rm j}^{1/2}+\pi^2 }\,.
\label{eq:phi}
\end{equation}
This approximate solution have the following asymptotical behaviour:
\begin{equation}
\Phi=
\begin{cases}
 \chi_{\rm j}; & \chi_{\rm j}\ll 1\\
\frac{\pi}{2}-\sqrt{\frac{1}{\chi_{\rm j}}}; & \chi_{\rm j} \gg 1.
\end{cases}
\label{eq:phiasim}
\end{equation}
Our approach yields a similar result to the analytical solution obtained by \cite{zdz15}, but is more general as it is also valid for large jet bending angles.

For bending angles $\Phi\lesssim\theta_{\rm j}$, the wind impact will trigger a weak recollimation shock or a sound wave in the jet, 
with a negligible impact on the general jet orientation and expansion. For $\Phi$ significantly larger than $\theta_{\rm j}$, 
the wind impact can trigger a strong recollimation shock in addition to a substantial deviation of the jet.
Bending can be thus considered relevant for the jet evolution if $\Phi>\theta_{\rm j}$. Since it is expected that $\theta_{\rm j}\ll 1$, one can fix
$\Phi=\theta_{\rm j}$ and adopt the solution for $\chi_{\rm j}\ll 1$ to obtain the following constraint for significant jet bending:
\begin{equation}
L_{\rm j}\lesssim \frac{(\gamma_{\rm j}-1)\dot{M_{\rm w}}\varv_{\rm w}c}{4\pi\gamma_{\rm j}\beta_{\rm j}}\approx 
2\times10^{36}\frac{\dot{M}_{\rm w,-7}\varv_{\rm w,8.5}(\gamma_{\rm j,0.3}-1)}{\gamma_{\rm j,0.3}\beta_{\rm j}}  \,{\rm erg~s}^{-1}\,.
\label{eq:ljphi}
\end{equation}

As the jet becomes bent by the wind impact an angle $\Phi$ from the initial jet direction and away from the star, the $y$-component of the bent jet velocity can be derived as $\approx\sin[\Phi] u$.

\subsection{Stellar wind impact: orbital effects}\label{orbef}

Orbital motion affects the propagation of a bent jet through the Coriolis effect. 
For small jet bending, i.e. $\Phi\lesssim \theta_{\rm j}$, the jet expansion eventually leads to 
a jet radius wider than the binary size, and the helical jet pattern gradually smooths out. 
Otherwise, if $\Phi$ is significantly larger than $\theta_{\rm j}$, the Coriolis effect should 
lead to a prominent helical structure of the jet trajectory (see Fig.~\ref{f2}). 

If the jet propagates in vacuum, 
the helical jet trajectory will be ballistic. The presence of the wind however implies that: (i) the Coriolis force can 
be enhanced by the wind ram pressure in the $\bf{e}_x$-direction in the jet flow co-rotating frame; and (ii) 
the shape of the helical jet on large scales is affected by the wind presence as well, as the jet, when forced to bend to become helical, 
pushes against the wind. We discuss in what follows (i), and consider (ii) in Sect.~\ref{exp}.

If situation (i) happens, then the distance $y_{\rm turn}$, covered by the jet in the $\bf{e}_y$-direction before turning towards the 
$\bf{e}_x$-direction because of the Coriolis effect, is significantly smaller than in the ballistic case. Otherwise, the Coriolis force enhancement due to the wind presence can be considered irrelevant. 

The ballistic value of $y_{\rm turn}$ is:
\begin{equation}
y_{\rm turn}^{\rm bal}=\frac{\sin[\Phi]u}{\omega_{\rm o}}\approx 3\times 10^{14} \frac{\sin[\Phi]\beta_{\rm j}}{\omega_{\rm o,-4}}\,{\rm cm}\,, 
\label{balturn}
\end{equation}
where $\omega_{\rm o,-4}=\omega_{\rm o}/10^{-4}\,{\rm s}$ is the orbital angular velocity of the compact object, with the normalization corresponding to a period of $\sim 1$~day for a circular orbit.

The non-ballistic value of $y_{\rm turn}$ can be roughly estimated assuming balance between the jet momentum flux, 
and the wind momentum flux in the co-rotating jet flow frame, i.e. in the $\bf{e}_x$-direction:
\begin{equation}
\frac{\dot{P}_{\rm j}}{2\theta_{\rm bj}(y_{\rm turn}^2+z^2)}
\approx 
\rho_{\rm w}\varv_\perp^2=
\frac{\dot{M}_{\rm w}}{4\pi\varv_{\rm w}[z^2+(y_{\rm turn}+\mu\,d)^2]}(y_{\rm turn}+\mu\,d)^2\omega_{\rm o}^2\, ,
\label{nonbal}
\end{equation}
where $\theta_{\rm bj}\lesssim \Phi$ is the half-opening angle of the bent jet, $\mu=M_*/(M_*+M_{\rm co})$, with $M_*$ being the stellar mass, and the wind momentum flux is computed in 
the $\bf{e}_x$-direction adopting the velocity of the co-rotating jet flow frame,
$\varv_\perp=(d+y_{\rm turn})\omega_{\rm o}$. Note that, as the momentum transfer takes place perpendicularly to the bent 
jet initial direction, the jet thrust has been divided by the lateral surface of the jet. 

For $y_{\rm turn}\gg d$, Eq.~(\ref{nonbal}) can be simplified to:
\begin{equation}
\dot{P}_{\rm j}\approx \dot{M}_{\rm w}
\frac{ \theta_{\rm bj} \omega_{\rm o}^2 y_{\rm turn}^2}{2\pi\varv_{\rm w}}=
\dot{P}_{\rm w}
\frac{ \theta_{\rm bj} \omega_{\rm o}^2 y_{\rm turn}^2}{2\pi\varv_{\rm w}^2}
\,.
\label{yvac1}
\end{equation}
Making the additional assumption that $\theta_{\rm bj}\sim \theta_{\rm j}$ (see Sect.~\ref{uncerdis}), and using Eq.~(\ref{eq:Bj}), one obtains:
\begin{equation}
y_{\rm turn}=
\frac{A_{\rm turn}}{\sqrt{2\chi_{\rm j}}}\frac{\varv_{\rm w}}{\omega_{\rm o}}\approx
2\times 10^{12} \frac{A_{\rm turn}}{\chi_{\rm j}^{1/2}} \frac{\varv_{\rm w,8.5}}{\omega_{\rm o,-4}}\,{\rm cm}\,,
\label{yvac2}
\end{equation}
where $A_{\rm turn}$ is a constant of the order of one, and accounts (as $\theta_{\rm bj}$) for the details of the wind-jet interaction, which are to be characterized through numerical simulations.

For the helical trajectory to be non-ballistic, it is required that $y_{\rm turn}<y_{\rm turn}^{\rm bal}$. 
Comparing Eqs.~(\ref{balturn}) and (\ref{yvac2}), and adopting for simplicity the approximation $\sin[\Phi]\sim\Phi\sim \chi_{\rm j}$, the stated condition implies:
\begin{equation}
\frac{2^{1/2}\chi_{\rm j}^{3/2}\beta_{\rm j}}{A_{\rm turn}}>\varv_{\rm w}/c\,,
\end{equation}
which can be rewritten as:
\begin{equation}
L_{\rm j}\lesssim 6\times10^{36} \frac{\theta_{\rm j,-1}\dot{M}_{\rm w,-7}\varv_{\rm w,8.5}^{1/3}(\gamma_{\rm j,0.3}-1)}{\gamma_{\rm j,0.3}\beta_{\rm j}^{1/3}} \mbox{ erg s}^{-1}\,.
\label{orbL}
\end{equation}
It is worth noting that, if $y_{\rm turn}<y_{\rm turn}^{\rm bal}$ is fulfilled 
and the jet were still supersonic when becoming helical, shocks could develop along the non-ballistic trajectory; note that the more relativistic the jet, the stronger the shocks will be \citep{bos12}.

\subsection{Stellar wind impact: large-scale evolution}\label{exp}

The jet flow moves through a channel made of shocked wind and shaped by jet bending plus orbital motion. When the helical trajectory is non-ballistic, the channel forces the jet flow to deviate and the whole helical structure becomes tighter. When the jet starts getting helical, the flow velocity in the $\bf{e}_y$-direction can be estimated from:
\begin{equation}
\varv_y\sim y_{\rm turn}\omega_{\rm o}\sim 2\times 10^8\, \frac{A_{\rm turn}}{\chi_{\rm j}^{1/2}} \varv_{\rm w,8.5} \,{\rm cm~s}^{-1}\ll u\,.
\label{varvy}
\end{equation}
Such a low value for $\varv_y$ implies that the fast jet flow should be deflected towards the $\bf{e}_z$- and $\bf{e}_x$-directions. In addition, while the jet turns, wind is also accumulated in the directions the jet does not temporarily point to. This implies that, as the jet pushes against the accumulated wind material in the $\bf{e}_z$-direction, the jet flow will slow down in that direction as well. It is therefore expected that, unless the jet flow itself is already strongly braked (or $\Phi$ rather small), the flow will move the fastest in the $\bf{e}_x$-direction; i.e. azimuthally with respect to the binary orbital axis, but in the direction opposite to the orbital motion.

It is hard to provide a more precise description of the helical jet geometry, or the jet flow velocity, as there are
several factors that cannot be captured in the present analysis. The hydrodynamics of the wind-jet interaction in the
helical structure is complex even assuming that the flows are laminar. In addition, although the jet flow velocity is
initially $u$, its value can be significantly lower if severe jet disruption, and/or wind mass-loading, and/or the
formation of a thick shear layer surrounding the jet, occur. All these three processes are related to instability
development, which is expected to happen in the present scenario: (i) strong growth of Kelvin-Helmholtz instabilities
already on the binary scales \citep[see][for numerical computations]{per08,per10,yoo15}; (ii) Rayleigh-Taylor
instabilities at the wind-jet interface along the helical jet structure \citep[see, e.g.,][]{mpmk14,imnwb15}. 

Severe
jet disruption can slow down and heat up the jet. On the other hand, pressure decreases away from the binary, and thus
the disrupted flow will tend to accelerate. However, strong wind-jet mixing is expected under jet disruption, which
will weaken this acceleration. Jet disruption and wind-jet mixing thus affect the jet helical structure, modulating its
dynamics and homogenizing it. Unfortunately, it is not known at which distance, or to which degree, the jet and the
wind flows mix and the whole structure homogenizes, although simulations of high-mass binaries with a non-accreting
pulsar \citep{bos12,bos15} suggest that this might already happen after just a few turns of the helical jet.

As explained, both the wind presence and orbital motion tighten the jet helical trajectory, but the jet reacts exerting a force on the surrounding wind, pushing it away from the binary. On the other hand, whereas on binary scales the wind transfers thrust to the jet in the $\bf{e}_y$-direction, on larger scales orbital motion redistributes this thrust, and the final half-opening angle of the jet helical structure should thus be $\sim\Phi$. As noted, the jet pushes wind when turning, hence transferring some momentum back to the wind, although one can consider this shocked/pushed wind surrounding the helical jet as part of the same structure. Far enough from the binary, wind-jet momentum transfer is negligible and the boundary of the helical structure and the unshocked wind motion become parallel (see Fig.~\ref{f2}), pointing radially with respect to the star.

Assuming maximum wind-jet mixing, the final velocity of the helical structure flow away from the binary can be estimated from:
\begin{equation}
\varv_{\infty}\sim \left(\frac{8\pi L_{\rm j}}{\Omega_{\rm hel}\dot{M}_{\rm w}}\right)^{1/2}\,,
\end{equation}
where $\Omega_{\rm hel}$ is the solid angle subtended by the helical jet, as seen from the star at large distances, which can be estimated using Eq.~(\ref{eq:phiasim}) as:
\begin{equation}
  \Omega_{\rm hel}\approx 
  \begin{cases}
    \pi \chi_{\rm j}^2 & \chi_{\rm j}\ll 1\\
    4\pi \chi_{\rm j}^{-1/2} & \chi_{\rm j}\gg 1\,.
  \end{cases}
\end{equation}
Note that the case $\chi_{\rm j}\gg 1$ corresponds to a bent jet pointing in the orbital plane (the relevant angle is then $\pi/2-\Phi$), similar to the case of a high-mass binary hosting a non-accreting pulsar.

For the case of a relatively small $\Phi$-value, one can write:
\begin{equation}
\varv_{\infty}\sim 
\left(\frac{2\theta_{\rm j}(\gamma_{\rm j}-1)\varv_{\rm w} c}{\pi\beta_{\rm j} \gamma_{\rm j} \chi_{\rm j}^{3}}\right)^{1/2}
\approx 5\times 10^8 \frac{\theta_{\rm j,-1}^{1/2}(\gamma_{\rm j,0.3}-1)^{1/2}\varv_{\rm w,8.5}^{1/2}}{\beta_{\rm j}^{1/2} \gamma_{\rm j,0.3}^{1/2}\,\chi_{\rm j}^{3/2}} \mbox{ cm s}^{-1}\,.
\label{vinf}
\end{equation}
This result shows that, on the largest scales, for small bending angles, the jet can eventually recover mildly relativistic velocities, but for moderate $\Phi$-values, say $\sim 0.3$, $\varv_\infty$ is well below $u$.  

\subsection{Stellar wind impact: weak-jet case}

It is worth deriving as well the conditions for the case of a weak jet:\\
(i) One way is to set the bending angle to $\Phi\gtrsim 1$, which yields:
 \begin{equation}
L_{\rm j}\lesssim \frac{\dot{M_{\rm w}}\varv_{\rm w}c\theta_{\rm j}(\gamma_{\rm j}-1)}{4\pi\gamma_{\rm j}\beta_{\rm j}}
\approx 
2\times10^{35} \frac{\dot{M}_{\rm w,-7}\varv_{\rm w,8.5}\theta_{\rm j,-1}(\gamma_{\rm j,0.3}-1)}{\gamma_{\rm j,0.3}\beta_{\rm j}}  \,{\rm erg~s}^{-1}\,.
\label{eq:ljbj1}
\end{equation}
(ii) A slightly different condition is obtained if one takes $\varv_{\infty}\sim\varv_{\rm w}$, i.e.: 
 \begin{equation}
\chi_{\rm j}\sim \left(\frac{\theta_{\rm j,-1}(\gamma_{\rm j}-1)}{\gamma_{\rm j}\beta_{\rm j}\varv_{\rm w, 8.5}}\right)^{1/3}\gtrsim 1
\end{equation}
yielding:
\begin{equation}
L_{\rm j}\lesssim 
2\times10^{35} \frac{\dot{M}_{\rm w,-7}\varv_{\rm w,8.5}^{4/3}\theta_{\rm j,-1}^{2/3}(\gamma_{\rm j,0.3}-1)^{2/3}}{\gamma_{\rm j,0.3}^{2/3}\beta_{\rm j}^{2/3}}  \,{\rm erg~s}^{-1}\,.
\label{eq:ljbj2}
\end{equation}
This second condition implies that the large-scale jet evolution is fully dominated by the interaction with the wind and the associated mass-load. 

If the jet has less power than these critical values, it will only affect a small volume of the binary surroundings at large distances. The termination regions of such a jet will be strongly affected by the stellar wind dynamics, and not the other way around \citep[e.g. the cases studied for instance in][]{bos11b}.

\section{Summary and discussion}
\label{disc}

\subsection{Results summary}

We conclude that, if the wind-induced bending angle of the jet is larger than its initial half-opening angle, i.e. $\Phi>\theta_{\rm j}$, and the conditions given in Eqs.~(\ref{eq:ljphi}) and (\ref{orbL}) are fulfilled, the jets of high-mass 
microquasars will suffer: (i) asymmetric recollimation shocks within the binary (see Eq.~\ref{recL}), (ii) bending away from the star (see Eqs.~\ref{eq:phi} and \ref{eq:ljphi}), (iii)
enhanced bending against orbital motion caused by the wind presence under the Coriolis effect (see Eqs.~\ref{yvac2} and \ref{orbL}), finally (iv) becoming a non-ballistic helical structure. Conclusions (i) and (ii) were already reached by numerical relativistic \citep{per08,per10,per12}, and numerical and analytical non-relativistic jet studies \citep{yoo15,zdz15}, although we generalize here the characterization of the jet bending angle $\Phi$. Conclusions (iii) and (iv) are for the first time discussed in detail here.
We also propose that
the recollimation shock, jet bending within the binary, and a further non-ballistic helical jet trajectory, lead to perturbations that will 
likely trigger instability growth that will partially or totally disrupt the jet, and can load it with wind material. These 
perturbations can also be associated to sites for particle acceleration (see Sect.~\ref{obssig}). Finally, on scales much larger 
than the binary, we conclude that the jet will likely become a non-relativistic, wind-loaded outflow, with a half-opening angle $\sim\Phi$ (see Eq.~\ref{vinf}). On those scales, 
the helical jet pattern, and a differentiated wind-jet structure, may be more or less washed out depending on the disruptive degree of 
instabilities within the binary system, and/or in the jet helical structure further up.

\subsection{Major sources of uncertainty}\label{uncerdis}

The values of the wind and the jet parameters, $\dot{M}_{\rm w}$, $\varv_{\rm w}$, $L_{\rm j}$, $\gamma_{\rm j}$, and $\theta_{\rm j}$, are by themselves 
uncertain, as it is hard to derive them from observations. Provided that they are known, the wind impact on the jet can 
be characterized by the angle $\Phi$, which depends on the parameter $\chi_{\rm j}$. This angle can be more precisely derived from simulations, although our prescription is consistent with numerical calculations \citep[see][with the caveat that these simulations were non-relativistic]{zdz15}. 
Additionally, there is the parameter $\theta_{\rm bj}$ (see Eq.~\ref{nonbal}), whose determination particularly requires simulations that have not been carried out yet. 
The chosen relation $\theta_{\rm bj}\sim\theta_{\rm j}$ when deriving Eq.~(\ref{yvac2}) is taken for illustrative purposes, although the $\theta_{\rm bj}$-value 
may easily be larger. Therefore, $\theta_{\rm bj}$ is possibly the most important source of 
uncertainty in the estimates derived in this work. This uncertainty ranges from a factor to order of one, if the jet is not significantly 
disrupted on the binary scales, up to an order of magnitude, for full jet disruption on the binary scales. Qualitatively, $\Phi$ and 
$\theta_{\rm bj}$ are expected to be positively correlated, but only simulations can tell how much. Finally, despite being less uncertain than $\theta_{\rm bj}$, the parameter that allows for the refinement of $y_{\rm turn}$ in Eq.~(\ref{yvac2}), $A_{\rm turn}$, should be also derived from simulations.

\subsection{Observational signatures}\label{obssig}

The periphery of the binary system may be the most suitable region for particle acceleration, and gamma-ray production,  being
also enough transparent to gamma rays above $\sim 10$~GeV \citep[e.g.][]{kha08,bos09}. In fact, the wind-jet interaction  may be
the powering mechanism of the gamma rays in the range $\sim 10^{-1}-10^4$~GeV observed from high-mass microquasars such  as
Cyg~X-1 and Cyg~X-3 \citep{alb07,tav09,abd09,sab10,mal13,bod14}. The most suitable regions for particle acceleration, and 
radiation, given the high stellar photon density on these scales, would be the jet recollimation shock. Further up, under the  
non-ballistic helical jet trajectory, shocks may occur and could accelerate particles, although at those distances energy losses
may be dominated already by adiabatic cooling and particle escape, given the different dependence with distance from  the binary
($r$) of radiative versus non-radiative losses, expected to be $\propto 1/r^2$ and $1/r$, respectively. This effect would yield lower radiation efficiencies the farther from the binary.

Radio emission, produced where the stellar wind is already transparent at low frequencies \citep[e.g.][]{szo07,bos09b}, far enough from the star, could be also detectable. A good candidate could be the region where the jet becomes a helical non-ballistic structure and beyond, where weak shocks and turbulence could occur leading to Fermi~I and stochastic particle acceleration \citep[see, e.g.][in the context of microquasar jets]{rie07}. Regarding radio signatures of wind-jet interaction, there is evidence on milliarcsecond (mas) scales of a sort of helical pattern in the jet of Cyg~X-3 \citep{mio01,mil04,tud07}, and possible jet bending in Cyg~X-1 \citep{sti01}. Unfortunately, we can only provide with approximate quantitative predictions for the jet properties of Cyg X-3 and Cyg~X-1 at the onset of the helical pattern, and for the flow velocity very far from the binary. At intermediate distances, numerical simulations accounting for radiation are required to characterize the actual aspect in radio of the helical jet flow, including the opening angle and other features of the radio structures. Nevertheless, one can still consider very approximately two extreme cases regarding the half-opening angle of $\lesssim 1^\circ$ of the jet of Cyg~X-1 found by \cite{sti01} (adopting the system inclination of \citealt{oro11} for the jet): (i) at the highest end of the possible jet power range \citep[say, up to $\sim 10^{38}$~erg~s$^{-1}$,][]{sel15}, the jet may be powerful enough for $\Phi\lesssim \theta_{\rm j}\sim 1^\circ$, given the stellar mass-loss rate in the source, $\sim 3\times 10^{-6}$~M$_\odot$~yr$^{-1}$ \citep[e.g.][]{hut76}; (ii) at more moderate but possible values of the jet power, say $\sim 10^{37}$~erg~s$^{-1}$, and assuming that the helical jet flow moves vertically with speed $\sim u\cos{[\Phi]}$, a constraint can be derived for $\varv_y<u\cos{[\Phi]}/\tan{[1^\circ]}\approx 5\times 10^8\beta_{\rm j}\cos{[\Phi]}$~cm~s$^{-1}$ on scales of $\sim 10$~mas (a de-projected linear size of $\sim 6\times 10^{14}$~cm).
This constraint on $\varv_y$ may be comparable to the value derived from Eq.~(\ref{varvy}). In fact, the wind and jet properties in Cyg~X-1 and Cyg~X-3 are such that the jets of these two high-mass microquasars could indeed suffer a recollimation shock within the binary system \citep[see the discussion in][and references therein]{zdz15}. In addition, although the relatively loose restrictions on jet power and wind mass-loss leave the value of $\Phi$ rather unconstrained, their winds are powerful and thus a non-ballistic nature for the jets seems likely. On the other hand, evidence of the presence of these jets on arcsecond \citep[e.g.][in the case of Cyg~X-3]{mar00}, and even arcminute \citep[e.g.][in the case of Cyg~X-1]{mar96,gal05}, scales implies that they do not fit, as expected, in the category of a weak jet dominated by the wind dynamics (see Eqs.~\ref{eq:ljbj1} and \ref{eq:ljbj2}). In any case, as noted, the high complexity of the process calls for thorough numerical studies for a proper comparison with observations, as the jet dynamics can be strongly non-linear on the multiple scales of interest, and the radio emitting flow may be just a part of the whole structure.

If one links the wind and the jet properties assuming an accretion scenario (e.g. Bondi spherical accretion), and a magnetically driven jet, one gets a relation between $L_{\rm j}$ and $\dot{M}_{\rm w}$, as done in \cite{bk12}. That work conclusions can be used to tentatively predict the value of $\Phi$. For instance, in the case of the powerful gamma-ray binary LS~5039, which hosts a compact object of unknown nature \citep[see the discussion on the nature of the compact object of LS~5039 in][and references therein]{bk12}, a putative jet may reach a power $\lesssim 10^{36}$~erg~s$^{-1}$. For a wind mass-loss $\sim 10^{-7}\,{\rm M}_\odot$~yr$^{-1}$, this could imply $\Phi\sim 0.1$, meaning that the jet will not be strongly bent. Nevertheless, strong bending cannot be fully discarded in this system, given the parameter uncertainties. It is noteworthy that radio images of this source showed morphology evolution along the orbit. This was attributed to the presence of a non-accreting pulsar in the system \citep[see][]{mol12}, although a bent jet affected by orbital motion cannot be discarded. In fact, a scenario with modest accretion, jet bending plus orbital motion, and some Doppler boosting (to reach the observed gamma-ray fluxes), may explain this source high-energy phenomenology, and is worthy of further study.

\subsection{Future prospects}

Although simulations of wind interaction with the jets of high-mass microquasars, in the relativistic \citep[e.g.][]{per08,per10,per12} and the non-relativistic 
\citep[e.g.][]{yoo15,zdz15} regimes, have been carried out on binary scales, the impact of orbital motion on the larger scale
evolution of these jets has not been studied yet. Here we provide with a semi-quantitative description of such a
scenario, although numerical computations of relativistic hydrodynamics, and eventually magnetohydrodynamics, are required to
better characterize the jet evolution, the regions where the jet suffers significant kinetic energy dissipation through shocks, the role on different spatial scales of the non-linear growth of instabilities
triggered at the wind-jet interfaces \citep[see, e.g.][in the context of pulsar-star wind interactions]{bos12,lam13,bos15}, both
on jet disruption and wind-jet mixing, etc. Finally, the magnetic field should be included as, even in the case of low-magnetization
plasmas, the field may be amplified in some regions, becoming dynamically important, through flow compression and braking. The
magnetic field could also play a role enhancing or suppressing instability growth, and in radiation. 

As a next step, and to check the semi-quantitative predictions done in this work, we plan to carry out 3-dimensional, 
relativistic hydrodynamical simulations including orbital motion, on scales significantly larger than the binary.  We note
nevertheless that the complexity of the flow evolution, and the different scales of interest (recollimation shock location, jet
bending, onset of the helical jet pattern, evolution of the helical structure, etc.), make such numerical calculations very
demanding.

\begin{figure}
\centering
\resizebox{\hsize}{!}{\includegraphics{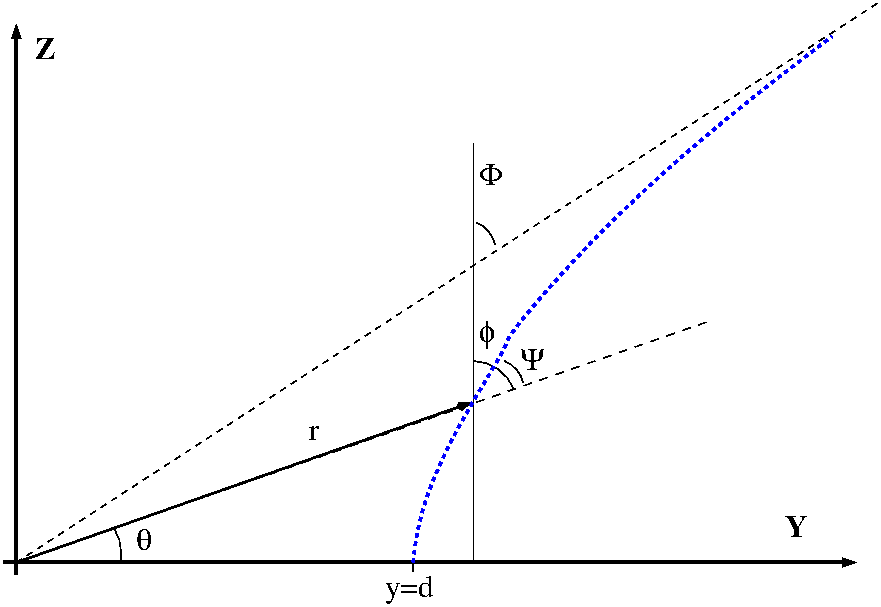}}
\caption{Sketch of the bent jet trajectory (blue dashed line) on the scales of the binary system under the impact of the stellar
wind.}
\label{af1}
\end{figure}

\begin{acknowledgements}
We thank the anonymous referee for his/her constructive and useful comments.
We want to thank Manel Perucho and Dmitry Khangulyan for fruitful discussions and comments.
V.B-R. acknowledges support by the Spanish Ministerio de Econom\'ia y Competitividad (MINECO) under grants
AYA2013-47447-C3-1-P, and MDM-2014-0369 of ICCUB (Unidad de Excelencia 'Mar\'ia de Maeztu').     
This research has been supported by the Marie Curie Career Integration Grant 321520.
V.B-R. also acknowledges financial support from MINECO and European Social Funds through a Ram\'on y Cajal fellowship.
BMV acknowledges partial  support  by  JSPS (Japan Society for the Promotion of Science):
No.16H00878, 2503786, 25610056, 26287056, 26800159. BMV also acknowledges MEXT (Ministry of Education, Culture, Sports, Science and Technology): No.26105521 and RFBR grant  12-02-01336-a.
\end{acknowledgements}

\bibliographystyle{aa}
\bibliography{ALLreferences}

\appendix

\section{Set of equations}
\label{soe}

The evolution of the conical jet bending angle with distance can be described through the 
following set of ordinary differential equations (ODE):
\begin{equation}
\begin{cases} 
\frac{d\Psi}{dt}=-\frac{d\theta}{dt}-\frac{d\phi}{dt}=
-\frac{\sin[\Psi]}{\sin[\phi+\Psi]}\frac{\beta_{\rm j}}{r}
-\frac{2\sin[\Psi]^2}{\cos[\phi]}\frac{\chi_{\rm j}}{r^2} \\ 
\frac{d\phi}{dt} = \frac{2\sin[\Psi]^2}{\cos[\phi]}\frac{\chi_{\rm j}}{r^2}\\ 
\frac{dr}{dt}=\beta_{\rm j}\cos[\Psi]
\end{cases}
\label{set}
\end{equation}
where
\begin{equation}
\chi_{\rm j}=\frac{\theta_{\rm j}(\gamma_{\rm j}-1)\dot{M}_{\rm w}\varv_{\rm w} c}{4\pi\beta_{\rm j} \gamma_{\rm j} L_{\rm,j} }\,,
\label{ap:Bj}
\end{equation}
$t$ and $r$ are non-dimensional time and length in units of $d/c$ and $d$, respectively, $\Psi$ is the angle between the direction from the normal star and the jet velocity, and $\phi$ is the angle between the jet velocity and the normal to the orbital plane (see Fig.~\ref{af1}).

We solve the set of ODE~(\ref{set}) numerically with the following initial conditions:
\begin{equation}
\begin{cases} 
\Psi(0)=\pi/2 \\ 
\phi(0) = 0 \\ 
r(0)=1.
\end{cases}
\label{init}
\end{equation}
We find then an approximated formula for the asymptotic solution of the angle $\phi$, where $\Phi=\phi(\infty)$: 
\begin{equation}
\Phi=\frac{\pi^2 \chi_{\rm j}}{2\pi \chi_{\rm j} +4\chi_{\rm j}^{1/2}+\pi^2 }.
\label{asimp}
\end{equation} 
The approximated formula fits the numerical solution with an accuracy better than 10\%. 

\end{document}